\documentclass[useAMS,usenatbib]{mn2e}

\usepackage{graphicx}
\usepackage{lscape}
\usepackage{amsmath}
\usepackage{textcomp}
\usepackage{longtable}
\usepackage[colorlinks,linkcolor=black,anchorcolor=black,citecolor=blue]{hyperref}

\title[A large sample of Am candidates]{A Large Sample of Am Candidates from LAMOST Data Release 1}
\author[Wen Hou et al. ]{Wen Hou$^{1,2}$, 
ALi Luo$^{1}$\thanks{email:lal@bao.ac.cn}, 
Haifeng Yang$^{1,2,3}$,
Peng Wei$^{1,2}$
Yongheng Zhao$^{1}$,
Fang Zuo$^{1}$,\and
Yihan Song$^{1}$,
Bing Du$^{1}$,
Zhongrui Bai$^{1}$,
Yong Zhang$^{4}$,
Yonghui Hou$^{4}$,
and Xiaowei Liu$^{5,6}$\\ \\
$^{1}$Key Laboratory of Optical Astronomy, National Astronomical Observatories, Chinese Academy of Sciences,
Beijing 100012, China\\
$^{2}$University of Chinese Academy of Sciences, Beijing 100049, China\\
$^{3}$School of Computer Science and Technology, Taiyuan University of Science and Technology, Taiyuan,030024, China\\
$^{4}$Nanjing Institute of Astronomical Optics and Technology, National Astronomical Observatories, Chinese Academy of Sciences,\\ Nanjing 210042, China\\
$^{5}$Kavli Institute for Astronomy and Astrophysics, Peking University, Beijing 100871, China\\
$^{6}$Department of Astronomy, Peking University, Beijing 100871, China\\}

\begin{document}


\pagerange{\pageref{firstpage}--\pageref{lastpage}} \pubyear{2014}

\maketitle

\label{firstpage}

\begin{abstract}
We present a sample of metallic-line star (Am) candidates from the Large sky Area Multi-Object fiber Spectroscopic Telescope Data Release one (LAMOST DR1). According to the characteristic of under-abundance of calcium and overabundance of iron element of Am stars, we propose an empirical separation curve derived from line indices of Ca II K--line and iron lines we choose for low resolution spectra. 3537 Am candidates are ultimately selected from more than 30,000 stars which are classified as A--type or early--F stars by both LAMOST pipeline and visual inspection. Then we make some analysis on this sample and finally provide a list of these Am candidates with 10 relevant parameters. Comparing with other catalogues, Am candidates selected from LAMOST DR1 are much fainter on the whole. Obviously, our list is an important complementary to already known bright Am catalogues, and it offers valuable material for the research on this type of chemically peculiar stars.
\end{abstract}


\section{Introduction}

Metallic-line stars (Am) which are firstly described by \citet{1940ApJ....92..256T}, are chemically peculiar A-- or early F-- type stars. Effective temperature of this group stars lies between 7000 K and 9000 K, or 10000 K if hot Am stars are included. They are mainly characterized by an apparent surface under--abundance of calcium or scandium and over--abundance of iron--group and heavier elements. In the past few decades, numerous studies have investigated on many aspects of this group of stars, such as the abundance analysis of individual element, rotational velocity, binarity, pulsation and diffusion of Am stars.

The abundance behaviour of individual element(Li, Ca, Fe, etc.) in the Am stars  has been studied in details in a number of papers \citep{1971A&A....11..325S, 1987MNRAS.226..361G, 1991A&A...249..205B, 2000A&AS..144..203H}. It has been widely accepted that there is deficiency of some light elements and enrichment of heavy elements in the atmosphere of Am stars. Much attention is also paid to exploring the physical mechanism about this interesting phenomenon of chemical separation in the envelope, beneath a superficial HI convection zone of Am stars \citep{1996Ap&SS.237...77S}. It has been recognized that the diffusion theory developed by \citet{1970ApJ...160..641M} can explain the chemical anomalies of Am stars \citep{1976ApJ...210..447M, 1978ApJ...223..920V, 1983ApJ...269..239M, 1996A&A...310..872A}. Besides  chemical abundances peculiarity, most Am stars have smaller rotational velocity than normal A--type stars. Roughly speaking, Am stars have rotational velocity {\it vsini\/} $<$ 100 km $s^{-1}$ and {\it vsini\/} $>$ 100 km $s^{-1}$ with a few possible exceptions \citep{1973ApJ...182..809A}. The effect of the rotational velocity on the metallicity of Am stars has also been investigated by \citet{1979A&A....74...38B}. The duplicity among Am stars is another important issue in research of Am stars. Lots of observational studies have shown that most Am stars belonged to close binary systems \citep{1961ApJS....6...37A,1985ApJS...59..229A}. It is the property of duplicity for most Am stars that leads to the fact that this group of stars have the relatively slower rotational velocity than normal A--type stars on the whole. On these characteristic of Am stars, \citet{1970PASP...82..781C} made a review and summarized their observational properties in some detail.

In addition, several catalogues related to Am stars have been published so far \citep{1969AJ.....74..375C, 1973A&AS...10..385H, 1979A&AS...38..449C, 1986A&AS...64...21H, 1992BICDS..40...19H, 1988A&AS...76..127R, 2009AA...498..961R}. These samples are quite valuable for researches on Am stars. Most Am stars in these available catalogues are bright stars with V magnitude smaller than 10 mag. In order to deeply understand these chemically peculiar stars, more observational data is still needed to address problems unanswered in Am stars.

In this paper, we present a large sample of Am star candidates mined out from LAMOST DR1. Here the data and sample selection are described in section 2. In section 3, we introduce the method of selecting the candidates of Am stars. A detailed analysis of results including spatial distribution, spectral type and V magnitude are presented in section 4. We also compare the sample with other catalogues and show the difference in this section. Besides, a list of Am candidates from LAMOST DR1 is provided and the parameters of objects are also given. Finally, a brief summary is provided in section 5.

\section[]{Sample Selection}
The Large Sky Area Multi--Object Fiber Spectroscopic Telescope (LAMOST, also called the Guo Shou Jing Telescope) is a special reflecting Schmidt telescope with an effective aperture of 3.6 -- 4.9 m and a field of view of 5$^{\circ}$. It is equipped with 4000 fibers, with a spectral resolution of R $\approx$ 1800 and the wavelengths ranging from 3800 $\mathrm{\AA}$ to 9000 $\mathrm{\AA}$  \citep{2012RAA....12.1197C, 2012RAA....12..723Z}. The LAMOST DR1, based on both a year of pilot survey from 2011 October to 2012 June and the first year of the general survey from 2012 September to 2013 June, contains more than 2 million spectra with limiting magnitude down to V $\sim$ 19.5 mag\citep{2012RAA....12.1243L}. In order to collect a large sample of Am candidates as accurate as possible, a dataset of stars that are classified as A stars or F0 stars with high signal-to-noise (S/N) of g--band (S/N $\geq$ 50) are chosen from LAMOST DR1. This initial dataset includes 38485 spectra getting rid of the bad ones (spectra of missing partial data). 

\section[]{Am Candidates Selection}

According to the abundance anomalies of Am stars, we use the properties of calcium and iron elements to distinct Am stars from normal A stars. Firstly, spectral lines whose line indices can represent the abundance of calcium and iron elements respectively should be picked out reasonably, especially iron lines. Taking characteristics of low--resolution spectra from LAMOST survey into account, we finally choose line indices of Ca II K--line and several groups iron lines to indicate abundance of calcium and iron respectively. When selecting iron lines, a few factors need to be considered such as intensity of lines. This part is introduced in detail in the following.

\subsection{Selection of iron lines list}
Because the relative error of measurement for the line index of a single iron line maybe quite large owing to its rather weak intensity in A--type stars, we have considered to choose several groups of iron lines for the calculation of iron line indices. The selection procedure consists of three steps.

Firstly, the wavelength region for selecting iron lines is limited within 4000 -- 5500 $\AA$. On the one hand, the intensity of metal lines in the red spectra ($\geq$ 6000 $\AA$) is almost very weak in spectra of A--type stars. On the other hand, the noise is relative large in the end of the blue spectra owing to its low response efficiency of the instrument. Secondly, we consider more than 20 iron lines with relatively stronger intensity on the basis of table III from \citet{1986A&AS...64..477C}. Finally, we get rid of several iron lines which are severely blended with lines of other elements, especially Ca, Sc. Meanwhile, in giant and supergiant of late--A stars, the abundance of Fe II and Ti II is supposed to increase according to the ionization equilibrium \citep{2009ssc..book.....G}. We are not sure whether this situation affects the selection of Am candidates. Therefore, lines blending with Fe II and Ti II elements are avoided as much as possible. Moreover, taking characteristic of iron lines in spectra with the resolution R$\approx$1800 into account, we map the rest lines to the spectra of LAMOST, and get 9 groups of iron lines which are listed in Table \ref{Fe list}. As shown in this table, the wavelength regions of line center for each group are given. Since the line center can not be well determined, therefore, we select two points as two ends of wavelength region for the line center. The left and the right points are used to calculate the line indices of iron lines, which is described in section 3.2.1.

\begin{table}
 \caption{9 groups of iron lines in vacuum wavelength.}
 \label{Fe list}
 \centering
 \begin{tabular}{@{}ccc}
  \hline
  	Number	 &\multicolumn{2}{c}{Wavelength region of line center} \\
   			 &  ~~~~left end($\AA$)  & ~~~~right end($\AA$) \\
  \hline
    1    &        4144.6                  &        4145.0 \\
    2    &        4199.5	              &        4205.2 \\
    3    &        4217.4	              &        4217.4 \\
    4    &        5234.4	              &        5235.4 \\
    5    &        5271.0	              &        5271.0 \\
    6    &        5371.0	              &        5373.0 \\
    7    &        5405.6	              &        5407.3 \\
    8    &        5425.6	              &        5436.0 \\
    9    &        5446.5	              &        5448.4 \\
  \hline
 \end{tabular}

\begin{flushleft}
{\sc Notes:}\\
Based on the fact that spectral lines of one element may be blended together in spectra of low resolution, each group contains several iron lines, rather than a single line.\\
\end{flushleft}
\end{table}

\subsection{Measurements of line indices}
Generally speaking, line index (or equivalent width) of a certain spectral line is calculated by integrating the continuum-normalized flux over the specified wavelength region of this line \citep{1998ApJS..116....1T, 2008AJ....136.2022L}. However, this method mentioned above is not suitable for calculating line index of iron lines in spectra of A--type stars due to its rather weak intensity as well as narrow feature bandpass. Therefore, it is essential for us to make some changes on the definition of the line index.

\begin{figure}
\center
 \includegraphics[width=84mm]{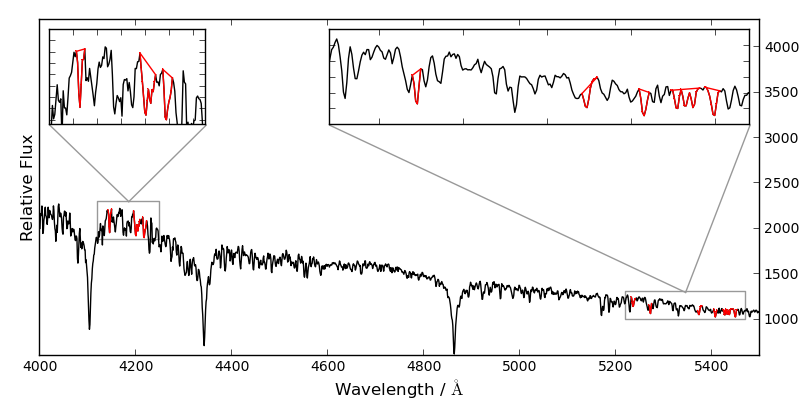}
 \caption{A spectrum from LAMOST DR1 with 9 groups of iron lines plotted in red line. Two amplified plots in this figure show the closed regions for calculating line index of the nine groups iron lines. }
 \label{Figure 1}
\end{figure}

Our method of line index measurement for iron lines, quite similar to the one proposed in the paper of \citet{2008AJ....136.2022L}, is described as follows. First of all, we transform the wavelength scale of spectra to rest frame by removing the radial velocity (RV). Here, RV is particularly represented by the mean value of two radial velocities which are derived from wavelength shifts of two strong Hydrogen absorption lines ($H\beta$ and $H\gamma$). The next step in calculation of iron lines index is how to get local pseudo-continuum. For each group of iron lines, a straight line is drawn through these two points of which the flux are the first peak values within 5 $\AA$ from either side the wavelength center respectively, where is different from the traditional method. Then, integration of the area surrounded by the straight lines and flux of the spectra is defined as line index of one group iron lines. The formula is defined as \[ EW = \int_{\lambda_1}^{\lambda_2}  (1-\frac{F_{I\lambda}}{F_{C\lambda}}) \, d\lambda \], where $\lambda_1$ and $\lambda_2$ are two wavelength limits of the local pseudo--continuum sideband. $F_{I\lambda}$ is the observed flux and $F_{C\lambda}$ represents the straight line drawn from the two points proposed above. Eventually, we add up line index of each group as the final line index of iron lines. In Figure \ref{Figure 1}, an example of a spectrum with 9 group of iron lines from LAMOST is shown. In order to see the integration region clearly, we enlarge the region for calculating the line index of iron in two amplified plots in Figure \ref{Figure 1}.

For consistency, the same method is applied to calculate line index of Ca II K--line except a smoothing process before finding the extreme points. Similarly, the radial velocity is determined by wavelength shift of K--line. For some spectra, noise appears in the profile of K line, which seriously interfere with finding the extreme points. Therefore, after removing the radial velocity, a gaussian smoothing is done on the K--line region to avoid the impact of noises, which is not needed in calculation of iron line index. Next we use the method proposed above to find the extreme points within 10 $\AA$ from the center of K--line, and compute the equivalent width of Ca II K--line by integrating the corresponding closed region of the observational spectra.

\subsection{Empirical line--index separation curve}
\begin{figure}
 \includegraphics[width=84mm]{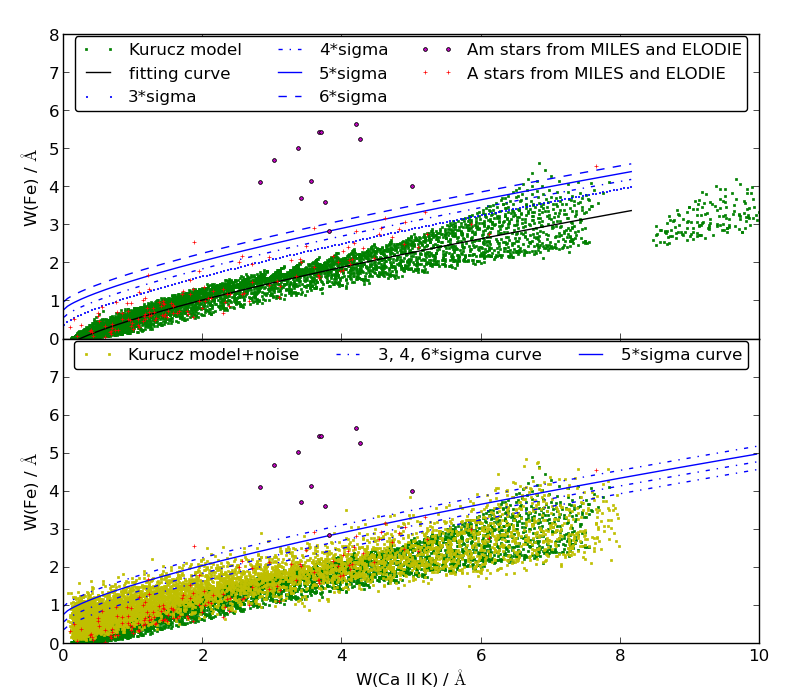}
 \caption{Ca II K--line vs. iron line index--index digram for spectra from Kurucz model and observations. Models from Kurucz with 7000 K $\leq T_\rmn{eff} \leq$ 10000 K, are plotted in green points. Models with the noise injected according to the S/N distribution of the sample from LAMOST are plotted in yellow points in the lower panel. 12 Am stars found in MILES and ELODIE library are represented by magenta filled circles, and 142 normal A stars by red plus, respectively. The fitting curve for the Kurucz models is plotted by the black solid line,  and the other four blue lines are considered as the candidates of the separation curve.}
 \label{Figure 2}
\end{figure}

Based on the line indices of Ca II K and iron, we plan to obtain an empirical separation curve in index--index diagram to distinguish Am stars from normal A--type stars. In order to find this curve , we need normal A--type and Am templates as much as possible. Based on this view, model templates from Kurucz and observational objects from spectral library are selected. Considering the fact that the effective temperatures of Am stars range from 7000 K to 10,000 K, a total of 6793 synthetic spectra are picked out from Kurucz model as templates of normal A stars. The parameters space covers the range 7000 K $\leq T_\rmn{eff} \leq$ 10000 K, in a step of 100 K below 7500 K and 250 K above 7500 K, 0.5 $\leq \rmn{log}g \leq$ 5.0, in a step of 0.25 dex, and -3.0 $\leq$ [Fe/H] $\leq$ 0.5, in a step of 0.2 dex below -1.0 and 0.1 dex above -1.0. Besides, 152 observational spectra from MILES and ELODIE library are also used, which include 140 normal A stars and 12 identified Am stars. All the spectra from Kurucz model and ELODIE library are supposed to convolve to the same resolution (R$\approx$1800) of spectra from LAMOST survey. 
Then we calculate the line indices of Ca II K--line and 9 groups of iron lines for model templates and the observational samples respectively. The distributions of line indices for both models and observational objects are shown in Figure \ref{Figure 2}.

Seen from the upper panel of Figure \ref{Figure 2}, there are a group of outliers plotted by green points in the right of this figure. This sample includes 258 models from Kurucz with 7000 K $\leq T_\rmn{eff} \leq$ 7500 K and 0.5 $\leq \rmn{log}g \leq$ 2.5. Basically, almost no Am stars are located in this region. Thus, these outliers are abandoned from Kurucz models in the selection of Am candidates. Moreover, there is another phenomenon in Figure \ref{Figure 2} that line indices of iron for observational normal A--type spectra (red plus) are slightly larger than those of model spectra (small green points) on the whole. It is probably attributed to the influence of noise.

The selection of separation curve by use of these two parts data containing Kurucz models and observational stars is described as follows. Firstly, a power function is chosen to fit these points from Kurucz models, and its curve is represented by a solid line shown in upper panel of Figure \ref{Figure 2}. Considering distribution characteristics of observational Am stars and normal A stars from both the Kurucz model and observational libraries in Figure \ref{Figure 2}, we select four lines which are 3$\sigma$, 4$\sigma$, 5$\sigma$ and 6$\sigma$ from the solid line respectively as candidates of the separation curve. Then, in order to pick out the most reasonable separation curve applied to the sample from LAMOST survey, white Gaussian noises are injected into the Kurucz spectra according to the S/N distribution of the sample from LAMOST shown in lower panel of Figure \ref{Figure 2}. A statistical results of applying the four candidate curves to distinguishing Am stars from normal A--type models with noises injected are listed in Table \ref{kurucz with noise}. From the first two rows, the misclassified percentages of normal A stars are too high to be accepted although all Am stars are separated correctly which are located above the 3$\sigma$ and 4$\sigma$ curves. For the 5$\sigma$ and 6$\sigma$ curves, the percentages of normal A stars mixed into Am are both acceptable, and one common Am star is missed. As shown in the Figure \ref{Figure 2}, this missed Am star is very close to the 5$\sigma$ curve. Considering that the spectra of Am templates we have obtained are rare, we believe there are quite a few Am stars located near the missed one. To guarantee the models representing normal A type stars below the separation curve and Am stars above this curve as much as possible, we finally choose the 5$\sigma$ line as the empirical line--index separation curve. 

\begin{table}
\center
 \caption{The number and ratio of kurucz models with noise and observational Am stars above each candidate curve.}
 \label{kurucz with noise}
 \begin{tabular}{@{}ccc}
  \hline
 candidate of      &        number of     &		percentage of \\
 separation curve   &        normal A $\mid$ Am  &      normal A $\mid$ Am ($\%$)\\
  \hline
  3$\sigma$			&		1407 $\mid$ 12		&		 20.71 $\mid$ 100.0\\
  4$\sigma$			&		~~509 $\mid$ 12		&		 7.493 $\mid$ 100.0\\
  5$\sigma$			&		~~147 $\mid$ 11		&		 2.164 $\mid$ 91.67\\
  6$\sigma$			&		~~~~29 $\mid$ 11		&	0.427 $\mid$ 91.67\\
  \hline
 \end{tabular}
 \begin{flushleft}
{\sc Notes:}\\
1. The total numbers of Kurucz models and Am stars are 6793 and 12 respectively.\\
2. The second column shows the number of normal A and Am above each candidate curve.\\
3. The values of the third column indicate the ratio of mixed normal A star to 6793 models and separated Am to 12 observational objects respectively.  
\end{flushleft}
\end{table}

\subsection{The Am candidates from LAMOST DR1}
\begin{figure}
 \includegraphics[width=84mm]{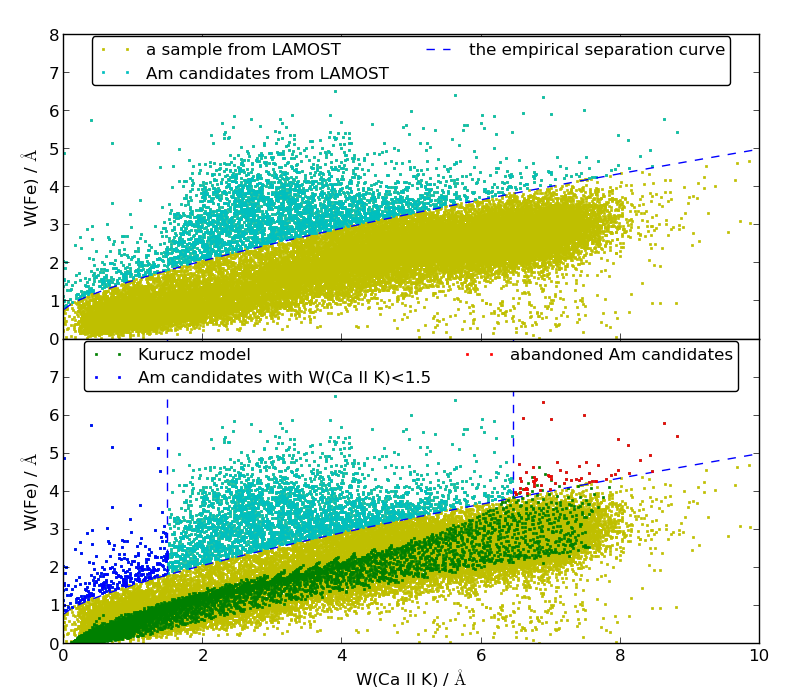}
 \caption{Ca II K--line vs. iron line index--index digram for the Am stars and normal A stars from LAMOST DR1. In the upper panel, the blue dashed line represents an empirical separation curve. Stars above the separation curve, represented by cyan points, are selected as initial Am candidates, and normal A stars of the sample is plotted in yellow points. In the lower panel, models from Kurucz are plotted in green points. Candidates with 1.5 $\leq$ W(Ca II K) $\leq$ 6.5 are represented by the cyan point, candidate with W(Ca II K) $<$ 1.5 by blue points and candidates with W(Ca II K) $>$ 6.5 by red points respectively.}
 \label{Figure 3}
\end{figure}

Applying the empirical separation curve developed in last subsection to the sample chosen from LAMOST DR1, more than 3000 stars are classified as Am stars candidates preliminarily, which is shown in the upper panel of Figure \ref{Figure 3}. Most of these candidates are concentrated in the area of 1.5 $\leq$ W(Ca II K) $\leq$ 6.5. They are relatively reliable since the 12 Am templates are all located in this area and the determinations of line indices are little affected by noises according to the test of Kurucz models with noises injected. There are 221 candidates accounting for less than 6\% in the area of W(Ca II K) $<$ 1.5, which is presented by blue points in the lower panel. Line indices of iron for these objects are rather small so that noises have a great effect on the calculation.  
The other part is 221 candidates whose line index of Ca II K is smaller than 1.5, which is presented by blue points in the same panel. From the test of Kurucz models with noises injected into mentioned above, we can make a conclusion that object with line index of Ca II K smaller than 1.5 are greatly affected by the noise. Therefore, this part of candidates with less reliability are provided in an individual list.
Additionally, we can see some Kurucz models with W(Ca II K) $>$ 6.5 marked by green points are mixed with Am candidates of this region from the lower panel. In order to make sure the sample as pure as possible, we remove the objects of LAMOST in this area from the whole sample of Am candidates. Finally, the rest 3316 objects of more reliable Am candidates are included in the major list we provide, which are plotted in cyan filled circles in the lower panel of Figure \ref{Figure 3}. 

\section{Result Analysis}

\subsection{Spatial distribution}
The spatial distribution of Am candidates and 38485 A and early--F stars from LAMOST DR1 is shown in Figure \ref{Figure 6}. The density of Galactic anti--center (GAC) is obviously higher than other areas. It can be explained by two factors: observational strategy and the real spatial distribution. On the one hand, since the Galactic anti--center survey is an import component of LAMOST project, more observations are carried out in this region. On the other hand, more objects are distributed in the Galactic disk actually. 

Although the spatial distributions are not uniform seen from Figure \ref{Figure 6}, the ratio of Am to A and early--F stars is basically consistent in different areas. Specifically, this ratio is 9.79\% for GAC and 7.58\% for non--GAC. The slight difference maybe resulted from the selection effect in GAC, which assigns a higher priority to A--type stars than F--type stars\citep{2014IAUS..298..310L}. The percentage of Am in all the A and early--F stars is relatively lower than the result provided by \citet{1996Ap&SS.237...77S}. Since the number of Am candidates from LAMOST DR1 depends on the selection criteria, the ratio may be higher if we relax our criteria.

\begin{figure}
 \includegraphics[width=84mm]{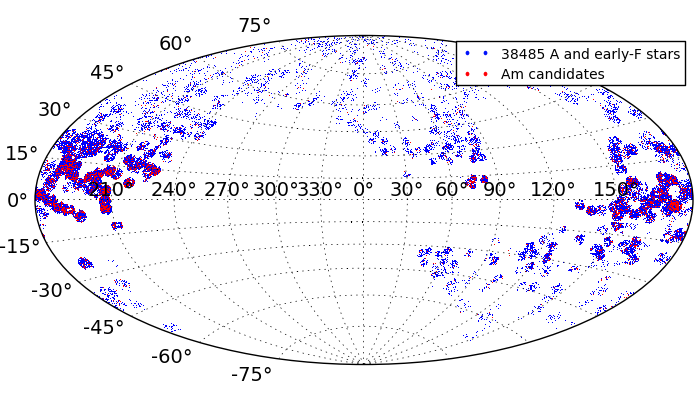}
 \caption{The distribution of objects from LAMOST DR1 in galactic coordinates. A sample of 38485 A and early--F type stars from LAMOST DR1 are represented by blue points, and 3537 Am candidates by red ones. {\color{red}{(In order to see clearly, the points in the legend box are artificially enlarged.) }}}
  \label{Figure 6}
\end{figure}

\subsection{Spectral type}

Due to the property of line strength anomalies, Am stars have three different spectral types on the basis of the intensity of the K, Balmer and metallic lines respectively. All these three spectral types of the 3537 Am candidates are obtained by template matching method. The 27 templates used in the matching process are selected from MILES library, and comprise of A--type and early F--type stars with spectral subtype from A0 to F3 (A8 is not found in MILES library), luminosity type from III to V. 

Based on the result of template matching, we find that the proportion of the candidates whose spectral type of hydrogen belong to mid-late A type stars is more than 80\% of all the Am candidates, which is consistent with the previous studies \citep{1971AJ.....76..896S,1996Ap&SS.237...77S}. We also find 3203 stars of the Am candidates of which K-line type is earlier than metallic-line type more or less. However, the rest stars are not satisfied with the definition of Am stars from the perspective of the discrepancy between K-line type and metallic-line type. It may be caused by the error of spectral type measurement or misclassification of normal A stars. Besides, according to the MK classification criteria, Am stars in which the Ca II K-line type is earlier than the metallic-line type by at least five spectral subtypes are classified as classical Am stars. And marginal Am stars are Am stars in which the difference between the K-line type and metallic-line type is less than five subclasses. A simple statistical result about the difference of spectral type between K-line type and metal-line type for Am candidates of LAMOST DR1 is shown in Table \ref{difference of spectral type}. From this table, the numbers of classical and marginal Am stars are comparable. We caution readers that the spectral types of some Am candidates maybe not very accurate without taking the uncertainty of measurement into account.

\begin{table}
\center
 \caption{The distribution for difference of spectral type between K-line type and metal-line type for Am candidates.}
 \label{difference of spectral type}
 \begin{tabular}{@{}lcc}
  \hline
 K type -- M type($\Delta$)                  &        Number    &		Percentage ($\%$) \\
  \hline
                  $\Delta$ $\geq$ 5          &        1422      &		 40.20\\
    0 $\textless$ $\Delta$ $\textless$ 5     &        1640      &		 46.37\\
                  $\Delta$ $\geq$ 0          &        475       &		 13.43\\
  \hline
 \end{tabular}
\end{table}

\subsection{Photometric analysis}
\subsubsection{Distribution of V magnitude}
\begin{figure}
 \includegraphics[width=84mm]{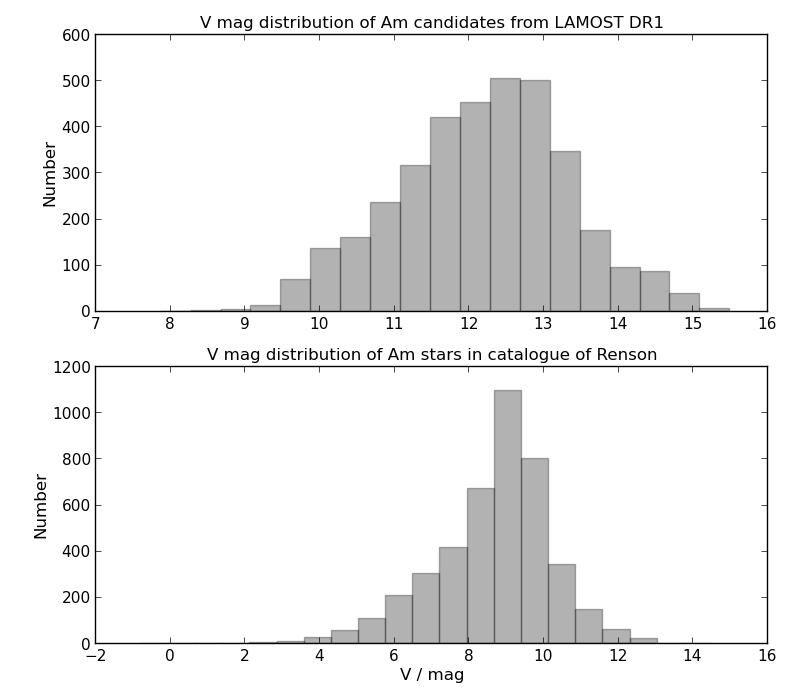}
 \caption{The upper panel shows the distribution of V magnitude for Am stars from LAMOST survey and it is concentrated between 12 and 13 mag. Am stars from Renson $\&$ Manfroid's catalogue is concentrated in 9 mag and almost all the stars are brighter than 12 mag, seen from the lower panel.
 }
  \label{Figure 5}
\end{figure}

In order to investigate the magnitude distribution of Am candidates from LAMOST DR1, we cross our list with UCAC4 \citep{2013AJ....145...44Z} and V magnitude is obtained for 3529 out of 3537 Am candidates through UCAC4. Objects from LAMOST DR1 are concentrated between 12 and 13 mag shown in the upper panel of Figure \ref{Figure 5}. 
Meanwhile, we compare our list with a catalogue of 4299 Am stars provided by \citet{2009AA...498..961R}, which is the latest and largest Am catalogue so far. 
It is found that only 18 Am stars are included in both the sample of LAMOST and the catalogue of Renson $\&$ Manfroid. V magnitude of Am stars from the catalogue of Renson $\&$ Manfroid is concentrated in 9 mag seen from lower panel of Figure\ref{Figure 5}. Conclusively, Am stars from LAMOST DR1 are much fainter on the whole, and they become an important complementary to already known bright Am catalogues. The sample offers valuable material for the research on this type of chemically peculiar stars.

\subsubsection{Light curve from LINEAR survey}

Am stars are likely to exhibit variability due to its binarity. Therefore light curves for a small portion of the Am candidates are also investigated in this paper. We obtain 187 light curves of the candidates in our list crossing with objects from LINEAR survey. Although it is difficult to detect the periodicity of the light curve owing to uneven sampling, we can see an obvious variability from the light curve of most of 187 candidates. A distribution of the light variability is shown in Figure \ref{Figure 6}. We can see that the amplitude of the variability is larger than 1 mag for more than half of the candidates from Figure \ref{Figure 6}. 

\begin{figure}
 \includegraphics[width=84mm]{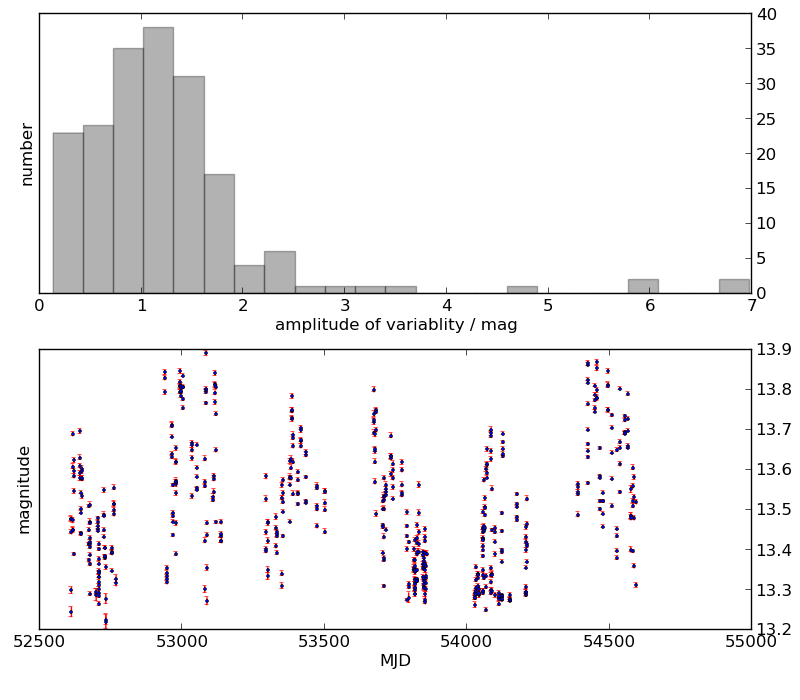}
 \caption{The upper panel presents the distribution of the amplitude of the light variability for  187 Am candidates. An example of light curve with the amplitude $\sim$ 0.8 mag is shown in the lower panel. The errorbar of magnitude is plotted in red line for each observation.}
  \label{Figure 6}
\end{figure}

\section{The list of Am candidates} 
A sample of Am candidates from LAMOST DR1 is provided in this section with 10 relative parameters. Table \ref{catalogue} shows the first 15 records of the major list\footnote{The full major list and another individual list including 221 less reliable candidates are provided in our package}. The column headings are Number, Designation, Ra, Dec, Confidence, V magnitude, K--line type, metallic--line type, logg and the error of logg respectively.

Confidence in the fifth column is a parameter defined to measure the degree of objects belonging to Am stars. The confidence equation is \[Confidence = k_1 \times \left[ \frac{D}{max(D)} \right]^{k_3} + k_2 \times \left[ \frac{\mu}{max(\mu)} \right]^{k_4} \], where D is introduced as the vertical distance between the object and the separation curve in index--index diagram, $\mu$ is the average inverse variance (average(1/$\sigma$)) of the flux in the region of Ca II K--line and iron line region, $k_1$, $k_2$ are two weight coefficients, and $k_3$, $k_4$ are coefficients reflecting the variation tendency of the confidence with D and $\mu$. Generally speaking, the confidence is greater with larger distance and higher $\mu$. Here, $\mu$ indicates the quality of particular region of the spectra like S/N. Considering D dominates the degree of objects belonging to Am, $k_1$ and $k_2$ are assigned to 0.8 and 0.2 empirically for candidates in the major list. Owing to a greater effect of noises on the confidence of lower reliable candidates in another list,  0.5 is assigned to $k_1$ and $k_2$. We also make an assumption that the confidence exhibits a non-linear growth with D or $\mu$. In other words, the value of the confidence is supposed to be larger with the increase of D or $\mu$, and it approaches 1 fast when D or $\mu$ exceeds a certain value. Based on these considerations, $k_3$ and $k_4$ in the equation are adopted as 0.25 empirically. Finally, in order to express the possibility of one object belonging to Am more clearly and formally, the range of confidence value is limited to [0, 1]. The possibility is low when the value is near 0, and high when near 1.

The photometric data of V magnitude is given in the sixth column, which is obtained by crossing with UCAC4. It is noted that for objects which are not found in UCAC4 or whose V magnitude is not provided in UCAC4, the photometric data is marked as `--' in the list. The next two columns show K--line and metallic-line types of Am candidates. We also provide logg and its error in the last two column. The parameters are obtained from LAMOST catalogue, in which atmospheric parameters are calculated by LASP pipeline (Bai. et al, in pre). Unluckily, there are 265 out of 3537 candidates whose parameters are unavailable in LAMOST catalogue, which are set to -999.

\begin{table*}
 \caption{The first 15 records of the major list: (1) the number; (2) the designation of target in LAMOST; (3,4) right ascension and declination; (5) confidence; (6) V magnitude; (7,8) K--line type and metallic--line type; (9,10) logg and its error. The full table is available online. }
 \label{catalogue}
 \begin{tabular}{@{}llcccccccc}
  \hline
Number & Designation & Ra{\color{red}{(deg)}} & Dec{\color{red}{(deg)}} & Confidence & V(mag) & K type & Metal type & logg(dex)& logg$\_$err(dex)\\
  \hline
1	&	J192220.85+430659.6		&	290.586875	&	43.116556	&	0.5296	&	12.033	&	A6	&	F3	&	3.765	&	0.343\\
2$^{*}$	&	J080110.99+031125.0		&	120.295820	&	3.1902940	&	0.7781	&	12.152	&	A2	&	F2	&	3.860	&	0.373\\
3	&	J191057.94+430242.6		&	287.741417	&	43.045194	&	0.7352	&	11.827	&	A4	&	F2	&	3.662	&	0.374\\
4	&	J193230.01+414020.9		&	293.125083	&	41.672494	&	0.4786	&	10.571	&	A7	&	F2	&	3.574	&	0.422\\
5	&	J162358.81+355039.8		&	245.995049	&	35.844410	&	0.7488	&	--		&	A3	&	F2	&	3.509	&	0.427\\
6$^{*}$	&	J054856.71+254910.6		&	87.2363120	&	25.819637	&	0.6088	&	11.415	&	A4	&	F3	&	3.803	&	0.425\\
7	&	J070347.29+225148.5		&	105.947050	&	22.863494	&	0.6824	&	13.401	&	A2	&	A9	&	3.938	&	0.383\\
8	&	J065904.34+250622.2		&	104.768100	&	25.106168	&	0.3728	&	13.381	&	A9	&	F3	&	3.822	&	0.398\\
9	&	J062832.82+032455.8		&	97.1367710	&	3.4155191	&	0.5530	&	11.924	&	A7	&	F3	&	4.029	&	0.362\\
10	&	J064845.75+203302.0		&	102.190660	&	20.550568	&	0.3790	&	12.395	&	A9	&	F2	&	3.974	&	0.357\\
11	&	J053122.37+404641.5		&	82.8432150	&	40.778216	&	0.7029	&	13.323	&	A3	&	F2	&	3.838	&	0.365\\
12	&	J063130.23+033022.0		&	97.8759780	&	3.5061371	&	0.5099	&	11.567	&	A3	&	F0	&	3.924	&	0.469\\
13	&	J052900.52+330703.7		&	182.252180	&	33.117695	&	0.6139	&	11.710	&	A2	&	F3	&	3.869	&	0.436\\
14	&	J064214.41+332925.2		&	100.560080	&	33.490360	&	0.5013	&	12.177	&	A6	&	F3	&	3.881	&	0.422\\
15$^{*}$	&	J084929.96-055014.5		&	132.374850	&	-5.837384	&	0.5516	&	9.751	&	A5	&	F0	&	3.752	&	0.328\\
  \hline
 \end{tabular}
 \begin{flushleft}
{\sc Notes:}\\
(1) The range of the confidence value in the fifth column is from 0 to 1. The possibility of one star belonging to Am is low when the value is near 0, and high for the value near 1. \\
(2) We give the spectral type from Ca II K--line and metallic lines in the sixth and seventh columns. From the perspective of MK classification, classical Am stars are the metallic-line stars whose difference between K--line type and metallic-line type is at least five subtypes, which is designated as Am. Marginal Am stars of which the difference of spectral type is less than five subtypes is designated as Am:.\\
(3) Objects that also contained in the catalogue of \citet{2009AA...498..961R} are marked a * in the top right corner of the number in the first column.\\
(4) The sample of Am candidates from LAMOST contains 3258 candidates ultimately. 243 objects in this list are observed twice or three times.
\end{flushleft}
\end{table*}

\section{Summary and Conclusion}
We present a sample of 3537 Am candidates from LAMOST DR1 containing 2 lists in this paper. The major list has 3316 candidates of higher reliability and the other individual list contains the rest 221 Am candidates of low reliability. By use of abundance anomalies of Am stars, an empirical separation curve is given in Ca II K--line vs iron line index diagram. When calculating the line index of spectral line, we make some changes on the basis of the method of measuring the equivalent width \citep{1998ApJS..116....1T}. This modified method is more suitable for calculating rather weak iron line of A--type stars as well as in spectra of low resolution. However, objects near the empirical separation curve maybe misclassified as Am stars because of some poor measurement for the noisy and low resolution spectra. Therefore a parameter of confidence is proposed to measure the degree of one candidate belonging to Am stars, which is shown in the fifth column of the final list.

Besides, some analysis are also made for the sample of Am candidates from LAMOST DR1 including spatial distribution, spectral type and photometry data. By the comparison of V magnitude between our list and Renson $\&$ Manfroid's catalogue, it is found that V magnitude of our list is $\sim$3 magnitude fainter than that of Renson $\&$ Manfroid's catalogue. In other words, the list from LAMOST DR1 offers much fainter candidates for the research of Am stars. Due to the limitation of spectra with low resolution, more sophisticated work can not be carried out such as a further identification of the candidates and analysis of individual elements. Hence spectra of high resolution or images from follow-up observations for our Am candidates are expected to carry out more in-depth studies on Am stars.

\section*{Acknowledgements}
We thank the anonymous referees for constructive comments. This work is supported by the National Key Basic Research Program of China (Grant No.2014CB845700), and the National Natural Science Foundation of China (Grant Nos 11390371, 11243004, 11303036).

Guoshoujing Telescope (the Large Sky Area Multi-Object Fiber Spectroscopic Telescope, LAMOST) is a National Major Scientific Project built by the Chinese Academy of Sciences. Funding for the project has been provided by the National Development and Reform Commission. LAMOST is operated and managed by the National Astronomical Observatories, Chinese Academy of Sciences.

\bibliography{reference}

\begin{thebibliography}{}

\bibitem[\protect\citeauthoryear{{Abt}}{{Abt}}{1961}]{1961ApJS....6...37A}
{Abt} H.~A.,  1961, ApJS, 6, 37

\bibitem[\protect\citeauthoryear{{Abt} \& {Levy}}{{Abt} \&
  {Levy}}{1985}]{1985ApJS...59..229A}
{Abt} H.~A.,  {Levy} S.~G.,  1985, ApJS, 59, 229

\bibitem[\protect\citeauthoryear{{Abt} \& {Moyd}}{{Abt} \&
  {Moyd}}{1973}]{1973ApJ...182..809A}
{Abt} H.~A.,  {Moyd} K.~I.,  1973, ApJ, 182, 809

\bibitem[\protect\citeauthoryear{{Alecian}}{{Alecian}}{1996}]{1996A&A...310..8%
72A}
{Alecian} G.,  1996, A\&A, 310, 872

\bibitem[\protect\citeauthoryear{{Burkhart}}{{Burkhart}}{1979}]{1979A&A....74.%
..38B}
{Burkhart} C.,  1979, A\&A, 74, 38

\bibitem[\protect\citeauthoryear{{Burkhart} \& {Coupry}}{{Burkhart} \&
  {Coupry}}{1991}]{1991A&A...249..205B}
{Burkhart} C.,  {Coupry} M.~F.,  1991, A\&A, 249, 205

\bibitem[\protect\citeauthoryear{Conti}{Conti}{1970}]{1970PASP...82..781C}
Conti P.~S.,  1970, PASP, 82, 781

\bibitem[\protect\citeauthoryear{{Coupry}, {vant Veer-Menneret} \&
  {Burkhart}}{{Coupry} et~al.}{1986}]{1986A&AS...64..477C}
{Coupry} M.~F.,  {vant Veer-Menneret} C.,    {Burkhart} C.,  1986, A\&AS, 64,
  477

\bibitem[\protect\citeauthoryear{{Cowley}, {Cowley}, {Jaschek} \&
  {Jaschek}}{{Cowley} et~al.}{1969}]{1969AJ.....74..375C}
{Cowley} A.,  {Cowley} C.,  {Jaschek} M.,    {Jaschek} C.,  1969, AJ, 74, 375

\bibitem[\protect\citeauthoryear{{Cui}, {Zhao}, {Chu}, {Li}, {Li}, {Zhang},
  {Su}, {Yao}, {Wang}, {Xing}, {Li}, {Zhu}, {Wang}, {Gu}, {Luo}, {Xu} \&
  {Zhang}}{{Cui} et~al.}{2012}]{2012RAA....12.1197C}
{Cui} X.-Q.,  {Zhao} Y.-H.,  {Chu} Y.-Q.,  {Li} G.-P.,  {Li} Q.,  {Zhang}
  L.-P.,  {Su} H.-J.,  {Yao} Z.-Q.,  {Wang} Y.-N.,  {Xing} X.-Z.,  {Li} X.-N.,
  {Zhu} Y.-T.,  {Wang} G.,  {Gu} B.-Z.,  {Luo} A.-L.,  {Xu} X.-Q.,    {Zhang}
  Z.-C.,  2012, Research in Astronomy and Astrophysics, 12, 1197

\bibitem[\protect\citeauthoryear{{Curchod} \& {Hauck}}{{Curchod} \&
  {Hauck}}{1979}]{1979A&AS...38..449C}
{Curchod} A.,  {Hauck} B.,  1979, A\&AS, 38, 449

\bibitem[\protect\citeauthoryear{{Gray} \& {Corbally} J.}{{Gray} \&
  {Corbally}}{2009}]{2009ssc..book.....G}
{Gray} R.~O.,  {Corbally} J. C.,  2009, {Stellar Spectral Classification}

\bibitem[\protect\citeauthoryear{{Guthrie}}{{Guthrie}}{1987}]{1987MNRAS.226..3%
61G}
{Guthrie} B.~N.~G.,  1987, MNRAS, 226, 361

\bibitem[\protect\citeauthoryear{{Hauck}}{{Hauck}}{1973}]{1973A&AS...10..385H}
{Hauck} B.,  1973, A\&AS, 10, 385

\bibitem[\protect\citeauthoryear{{Hauck}}{{Hauck}}{1986}]{1986A&AS...64...21H}
{Hauck} B.,  1986, A\&AS, 64, 21

\bibitem[\protect\citeauthoryear{{Hauck}}{{Hauck}}{1992}]{1992BICDS..40...19H}
{Hauck} B.,  1992, Bulletin d'Information du Centre de Donnees Stellaires, 40,
  19

\bibitem[\protect\citeauthoryear{{Hui-Bon-Hoa}}{{Hui-Bon-Hoa}}{2000}]{2000A&AS%
..144..203H}
{Hui-Bon-Hoa} A.,  2000, A\&AS, 144, 203

\bibitem[\protect\citeauthoryear{{Lee}, {Beers}, {Sivarani}, {Allende Prieto},
  {Koesterke}, {Wilhelm}, {Re Fiorentin}, {Bailer-Jones}, {Norris}, {Rockosi},
  {Yanny}, {Newberg}, {Covey}, {Zhang} \& {Luo}}{{Lee}
  et~al.}{2008}]{2008AJ....136.2022L}
{Lee} Y.~S.,  {Beers} T.~C.,  {Sivarani} T.,  {Allende Prieto} C.,  {Koesterke}
  L.,  {Wilhelm} R.,  {Re Fiorentin} P.,  {Bailer-Jones} C.~A.~L.,  {Norris}
  J.~E.,  {Rockosi} C.~M.,  {Yanny} B.,  {Newberg} H.~J.,  {Covey} K.~R.,
  {Zhang} H.-T.,    {Luo} A.-L.,  2008, AJ, 136, 2022

\bibitem[\protect\citeauthoryear{{Liu}, {Yuan}, {Huo}, {Deng}, {Hou}, {Zhao},
  {Zhao}, {Shi}, {Luo}, {Xiang}, {Zhang}, {Huang} \& {Zhang}}{{Liu}
  et~al.}{2014}]{2014IAUS..298..310L}
{Liu} X.-W.,  {Yuan} H.-B.,  {Huo} Z.-Y.,  {Deng} L.-C.,  {Hou} J.-L.,  {Zhao}
  Y.-H.,  {Zhao} G.,  {Shi} J.-R.,  {Luo} A.-L.,  {Xiang} M.-S.,  {Zhang}
  H.-H.,  {Huang} Y.,    {Zhang} H.-W.,  2014, in {Feltzing} S.,  {Zhao} G.,
  {Walton} N.~A.,   {Whitelock} P.,  eds, IAU Symposium Vol.~298 of IAU
  Symposium, {LSS-GAC - A LAMOST Spectroscopic Survey of the Galactic
  Anti-center}.
pp 310--321

\bibitem[\protect\citeauthoryear{{Luo}, {Zhang}, {Zhao}, {Zhao}, {Cui} \&
  {Li}}{{Luo} et~al.}{2012}]{2012RAA....12.1243L}
{Luo} A.-L.,  {Zhang} H.-T.,  {Zhao} Y.-H.,  {Zhao} G.,  {Cui} X.-Q.,    {Li}
  G.-P.,  2012, Research in Astronomy and Astrophysics, 12, 1243

\bibitem[\protect\citeauthoryear{{Michaud}}{{Michaud}}{1970}]{1970ApJ...160..6%
41M}
{Michaud} G.,  1970, ApJ, 160, 641

\bibitem[\protect\citeauthoryear{{Michaud}, {Charland}, {Vauclair} \&
  {Vauclair}}{{Michaud} et~al.}{1976}]{1976ApJ...210..447M}
{Michaud} G.,  {Charland} Y.,  {Vauclair} S.,    {Vauclair} G.,  1976, ApJ,
  210, 447

\bibitem[\protect\citeauthoryear{{Michaud}, {Tarasick}, {Charland} \&
  {Pelletier}}{{Michaud} et~al.}{1983}]{1983ApJ...269..239M}
{Michaud} G.,  {Tarasick} D.,  {Charland} Y.,    {Pelletier} C.,  1983, ApJ,
  269, 239

\bibitem[\protect\citeauthoryear{{Renson}}{{Renson}}{1988}]{1988A&AS...76..127%
R}
{Renson} P.,  1988, A\&AS, 76, 127

\bibitem[\protect\citeauthoryear{{Renson} \& {Manfroid}}{{Renson} \&
  {Manfroid}}{2009}]{2009AA...498..961R}
{Renson} P.,  {Manfroid} J.,  2009, A\&A, 498, 961

\bibitem[\protect\citeauthoryear{{Smith}}{{Smith}}{1996}]{1996Ap&SS.237...77S}
{Smith} K.~C.,  1996, Ap\&SS, 237, 77

\bibitem[\protect\citeauthoryear{{Smith}}{{Smith}}{1971a}]{1971A&A....11..325S}
{Smith} M.~A.,  1971a, A\&A, 11, 325

\bibitem[\protect\citeauthoryear{{Smith}}{{Smith}}{1971b}]{1971AJ.....76..896S}
{Smith} M.~A.,  1971b, AJ, 76, 896

\bibitem[\protect\citeauthoryear{{Titus} \& {Morgan}}{{Titus} \&
  {Morgan}}{1940}]{1940ApJ....92..256T}
{Titus} J.,  {Morgan} W.~W.,  1940, ApJ, 92, 256

\bibitem[\protect\citeauthoryear{{Trager}, {Worthey}, {Faber}, {Burstein} \&
  {Gonz{\'a}lez}}{{Trager} et~al.}{1998}]{1998ApJS..116....1T}
{Trager} S.~C.,  {Worthey} G.,  {Faber} S.~M.,  {Burstein} D.,
  {Gonz{\'a}lez} J.~J.,  1998, ApJS, 116, 1

\bibitem[\protect\citeauthoryear{{Vauclair}, {Vauclair} \&
  {Michaud}}{{Vauclair} et~al.}{1978}]{1978ApJ...223..920V}
{Vauclair} G.,  {Vauclair} S.,    {Michaud} G.,  1978, ApJ, 223, 920

\bibitem[\protect\citeauthoryear{{Zacharias}, {Finch}, {Girard}, {Henden},
  {Bartlett}, {Monet} \& {Zacharias}}{{Zacharias}
  et~al.}{2013}]{2013AJ....145...44Z}
{Zacharias} N.,  {Finch} C.~T.,  {Girard} T.~M.,  {Henden} A.,  {Bartlett}
  J.~L.,  {Monet} D.~G.,    {Zacharias} M.~I.,  2013, AJ, 145, 44

\bibitem[\protect\citeauthoryear{{Zhao}, {Zhao}, {Chu}, {Jing} \&
  {Deng}}{{Zhao} et~al.}{2012}]{2012RAA....12..723Z}
{Zhao} G.,  {Zhao} Y.-H.,  {Chu} Y.-Q.,  {Jing} Y.-P.,    {Deng} L.-C.,  2012,
  Research in Astronomy and Astrophysics, 12, 723

\end{thebibliography}

\bsp

\label{lastpage}

\end{document}